\documentstyle[aps,prb,multicol]{revtex}

\begin{document}
\input{epsf}
\draft
\preprint{}
\title{Analysis of the temperature-dependent quantum point contact 
conductance in view of the metal-insulator transition in two dimensions}
\author{V. Senz, T. Heinzel, T. Ihn, S. Lindemann, R. Held, and K. Ensslin}
\address{Solid State Physics Laboratory, ETH Z\"{u}rich, 8093
Z\"{u}rich,  Switzerland\\}
\author{W. Wegscheider}
\address{Walter Schottky Institut, TU M\"unchen, 85748 Garching, 
Germany, and\\
Institut f\"ur Angewandte und Experimentelle Physik, Universit\"at 
Regensburg, 93040 Regensburg, Germany\\}
\author{M. Bichler}
\address{Walter Schottky Institut, TU M\"unchen, 85748 Garching, 
Germany\\}
\date{\today}
\maketitle
\begin{abstract}
The temperature dependence of the conductance of a quantum 
point contact has been measured. The conductance as a function of the 
Fermi energy shows temperature-independent fixed points, located at roughly 
multiple integers of $e^{2}/h$. Around the first fixed point at e$^{2}$/h, the experimental data for 
different temperatures can been scaled onto a single curve. For pure thermal 
smearing of the conductance steps, a scaling parameter of one is expected.
The measured scaling parameter, however, is significantly larger than 
1. The deviations are interpreted as a signature of the 
potential landscape of the quantum point contact, and of the 
source-drain bias voltage. We relate our 
results phenomenologically to the metal-insulator transition in two 
dimensions.
\end{abstract}
\begin{multicols}{2}
Conductance quantization in short, quasi one-dimensional wires  
\cite{vanWees88,Wharam88} (``quantum point contacts''-QPCs) has been widely investigated since 
its discovery. Meanwhile, QPCs have become a key device for 
transport experiments in low-dimensional systems.\cite{Houten92}
Recently, QPCs have experienced renewed interest in relation to the 
metal-insulator transition (MIT) in two dimensions\cite{Altshuler2000}: 
it has been suggested that near the fixed point of the MIT, which 
typically occurs at a conductance of $G\approx 
e^{2}/h$, the carrier gas segregates into conductive 
puddles\cite{Eytan1998}, which are connected via quantum point contacts.\cite{Meir99} 
Within such a percolation picture, a statistical ensemble of QPCs determines the 
transport properties of the inhomogeneous carrier system, since the 
applied external voltage drops predominantly across the QPCs. Of 
particular importance is the fixed point in the temperature-dependent 
conductance of QPCs at $e^{2}/h$, which enters in the more general 
discussion of the quantum percolation model of Ref. \onlinecite{Meir99} 
and, within this model, corresponds to the fixed point of the metal-insulator transition in the 
two-dimensional carrier gas.\\
This is our motivation to study in detail the temperature dependence of the conductance of a 
single QPC, as the elementary unit of such a quantum percolation 
network, around this fixed point. 
Theoretically, one expects that the conductance $G(\mu ,T)$ of a 
spin-degenerate QPC as a function of the 
chemical potential $\mu$ and the temperature T, is independent of temperature at $G=ne^{2}/h$, 
with $n$ being an integer. These fixed points separate regions 
$(2n-1)e^{2}/h<G<2ne^{2}/h$, in which $G$ decreases as T is 
increased, from regions $2ne^{2}/h<G<(2n+1)e^{2}/h$, where 
$G$ increases as T is increased.\\
In a simple picture, $G(\mu,T=0)$ can be modelled by a sum of step functions.
For the fixed point lowest in energy, $n=1$, we obtain
\end{multicols}

\begin{equation}
    G(\mu,T)=\frac{2e^{2}}{h}\int dE\Theta(E-E_{1})\lbrack 
    -\frac{\partial f(E)}{\partial E}\rbrack dE = 
    \frac{2e^{2}}{h}\lbrack 1+e^{(E_{1}-\mu)/k_{B}T}\rbrack^{-1}. 
\end{equation}
\begin{multicols}{2}

Here, $f$ is the Fermi 
function, and $E_{1}$ is the energy of the first one-dimensional subband. Furthermore, 
we have assumed that the voltage drop across the QPC is small compared 
to $k_{B}T$. Since within this model, $G(\mu,T)$ has the shape of a 
Fermi function, the temperature-dependent conductance 
traces obtained for fixed $\mu$ can be scaled onto a single curve 
$G^{*}(\delta/T^{1/\alpha})$, with $\delta = (\mu-E_{1})/E_{1}$ and 
the scaling parameter $\alpha$ = 1. $G^{*}$ consists of  
two branches, corresponding to the metallic and the insulating 
region. Note that here, in contrast to the heavily debated scaling 
properties of two-dimensional systems,\cite{Altshuler2000} scaling does \textit{not} 
imply a possible quantum phase transition at T=0. Nevertheless, it is 
interesting in view of Ref. \onlinecite{Meir99} to compare $\alpha$ 
obtained here with the scaling parameters found in two-dimensional 
systems showing a MIT.\\
The temperature dependence of the conductance quantization 
in QPCs has been addressed in several 
experiments,\cite{Wees1989,Frost93,Dzurak93,Taboryski95,Yacobi1996,Thomas96,Thomas98} in particular with 
respect to the so-called ``0.7-feature''.\cite{Thomas96,Thomas98}
To our knowledge, the fixed point at $G=e^{2}/h$ 
and the temperature dependence around it has not 
been investigated yet. However, fixed points have been detected occasionally, while the observed 
behaviors around these fixed points 
vary greatly and definitely differ from the expectation expressed by 
eq. (1). Yacobi et al., \cite{Yacobi1996}, for example, observe a fixed 
point at $G=1.5e^{2}/h$ only, around which the metallic and the insulating 
phases appear reversed. This anomaly probably originates in the 
observed  non-universal conductance quantization. Thomas et al.\cite{Thomas96}, observe fixed points at  
$G=3e^{2}/h$ and $G=2e^{2}/h$, but not at $G=e^{2}/h$, which is 
possibly destroyed by the 0.7-feature.\\   
In the present paper, we report the observation of a series of clear fixed 
points (n=1\ldots 5) in a  QPC. We investigate $G(\mu ,T)$ around the fixed point at 
G $\approx$ e$^{2}$/h in further detail. $G(\mu,T)$ 
can be scaled reasonably well onto a single curve. The scaling 
parameter $\alpha$, however, is significantly larger than 1 (the value 
assumed in Ref. \onlinecite{Meir99}), and increases monotonically as the source-drain 
bias voltage is increased. Possible implications for describing the 
MIT in 2 dimensions in terms of a quantum percolation model are 
discussed.   
 \begin{figure}
\centerline{\epsfxsize=7.5cm \epsfbox{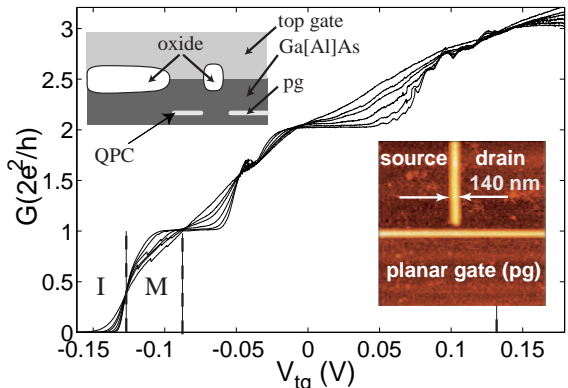}}
\caption{The lower right inset shows the surface topography of the sample, measured with an 
AFM directly after the oxidation step. The QPC
is defined by two oxide lines (bright lines), and connects source 
with drain.  
Upper left inset: cross section through the device, after deposition of the top 
gate, along a vertical line through the QPC in the lower inset. Below the 
oxide lines, the 2DEG is depleted. Main figure: conductance G as a function of the top gate voltage, 
taken at temperatures of 1.7K, 5K, 8K, 12K, 15K, and 20K. The 
metallic and insulating regions around $G=e^{2}/h$ are indicated by ``M'' and ``I''. 
Here, the planar gate voltage was -100mV, and $V_{b}$ =50 $\mu$V.}
\label{Fig1}
\end{figure}

The sample is a Ga[Al]As heterostructure, grown by molecular beam 
epitaxy, with the 2DEG $34$ nm below the surface. The 
 QPC is defined by local oxidation with an atomic force 
 microscope \cite{Held98} (insets in Fig.1). The 
 lithographic channel width is 80 nm, and its length is given by the 
 width of the oxide lines, i.e., 140 nm. The QPC  can be 
 tuned by voltages applied to a homogeneous top gate (tg) and to the planar 
 gate (pg). A top gate voltage changes $\mu$ throughout the sample. 
 However, G is still determined by the QPC, as long as the 
 surrounding 2DEG is not depleted.
 The sample is mounted in a $^{4}$He cryostat, in which the 
 temperature can be varied between 1.7 K and room temperature, or in the mixing chamber of a
 $^{3}$He/$^{4}$He - dilution refrigerator with a base temperature of
 90\,mK, respectively. The electron density of the ungated 2DEG 
 is 5.5$\cdot$10$^{15}$m$^{-2}$, and the mobility of the 2DEG is $93$ 
 m$^{2}$/Vs at T = 4.2 K.\\
 AC resistance measurements are performed in a four-terminal setup.
As the top gate voltage $V_{tg}$
or the temperature is changed, the voltage drop V$_{b}$ across the QPC is kept
constant by a feedback loop which adjusts the current accordingly. 
This constant voltage drop is important, since V$_{b}$ modifies 
$G(\mu,T)$, as will be discussed below. Figure 1 shows
G as a function of $V_{tg}$ for different temperatures. 
The conductance quantization has a characteristic temperature 
of $\approx$ 15K, which corresponds to an energy separation between the 
one-dimensional subbands of $\hbar \omega_{y}$ = 3.52 $k_{B}T \approx$ 4.5 meV. 
Here, $\omega_{y}$ denotes, within a parabolic approximation, the confining strength transverse to the 
transport direction. Fixed points occur at roughly multiple 
integers of e$^{2}$/h, and separate metallic regions from insulating 
regions.
The traces contain additional conductance fluctuations between 
the plateaus, which depend on the cooldown cycle, but are always 
present and perfectly reproducible within one cooldown. Probably, 
they are due to imperfect 
adiabatic coupling between the QPC and the reservoirs, or due to 
nearby scatters. At temperatures below 1K, the 
shapes of the steps are independent of temperature (not shown), 
indicating that the smearing of the steps is not solely determined by 
temperature.\\
We therefore model the QPC potential by a parabolic saddle 
point potential,\cite{Buttiker90} i.e. 
$V(x,y)=V_{0}-\frac{1}{2}m^{*}\omega_{x}x^{2}+\frac{1}{2}m^{*}\omega_{y}^{2}y^{2}$;
V$_{0}$ is the potential energy of the saddle point, m$^{*}$ the 
effective electron mass, and  $\omega_{x}$ denotes the curvature of 
the saddle point potential in transport direction. Using this model 
is justified since in the following, we focus on the regime G $<$ 
2e$^{2}$/h.\\ 
The transmission T($\epsilon$) for the lowest one-dimensional subband 
can be written as\cite{Buttiker90}
\begin{equation}
T (\epsilon)=\lbrack 1+e^{-\pi \epsilon} \rbrack^{-1},\epsilon=
2(E-\frac{1}{2}\hbar\omega_{y}-V_{0})/\hbar\omega_{x}.
\end{equation}
Here, $E$ denotes the energy of the incident electrons. 
Since the separation of the conductance plateaus corresponds to 
$\hbar \omega_{y}$ in energy, we can estimate the lever arm inside the QPC and 
in the regime of 
the first conductance step to $\frac{dE_{F}}{dV_{tg}}\approx$ 50 
meV/V, which we use to transform V$_{tg}$ into energy. 
In the limit of $V_{b}$ = 0 and for negligible temperatures, the width of the 
step edge corresponds 
to $\hbar \omega_{x}$. Hence, we can estimate $\hbar 
\omega_{x}\approx$ 2 meV from Fig. 1.\\
We proceed by studying the ``metal-insulator transition'' around 
$G=e^{2}/h$ in more detail (Fig. 2).
As $V_{b}$ is increased, the fixed point remains well-defined and 
shifts to smaller conductances. In addition, the step edge broadens 
significantly. In the insets of Figs. 2a and 2b, we show the best 
scaling achieved with the scaling variable $|\delta|/T^{\alpha}$, 
using $\alpha$ as a parameter. Scaling of the experimental data 
works well, although worse than for many  
two-dimensional systems showing an MIT.\cite{Senz2000} The scaling parameter $\alpha$, 
however, is larger than 1 and increases as $V_{b}$ is increased. We 
estimate the error of $\alpha$ to $\delta \alpha \approx \pm 0.3$ at 
small $V_{b}$, which increases for higher bias voltages. Thus, 
the MIT is clearly \textit{not} solely determined by thermal smearing, 
as eq.(1) suggests.\\
In order to understand the origin of the step edge slope and the experimental scaling 
parameter in more detail, we consider a simple model, which includes 
thermal smearing, the potential landscape in 
transport direction, described by $\omega_{x}$, and the bias voltage 
$V_{b}$. We therefore follow Ref. 8 and model the current $I$ by
\end{multicols}

\begin{equation}
    I=\frac{e}{\pi\hbar}\int T (\epsilon)\cdot 
    \lbrack f(E-(E_{F}+eV_{b}),T)-f(E-E_{F},T)\rbrack dE,
\end{equation}
 \begin{figure}
\centerline{\epsfxsize=12cm \epsfbox{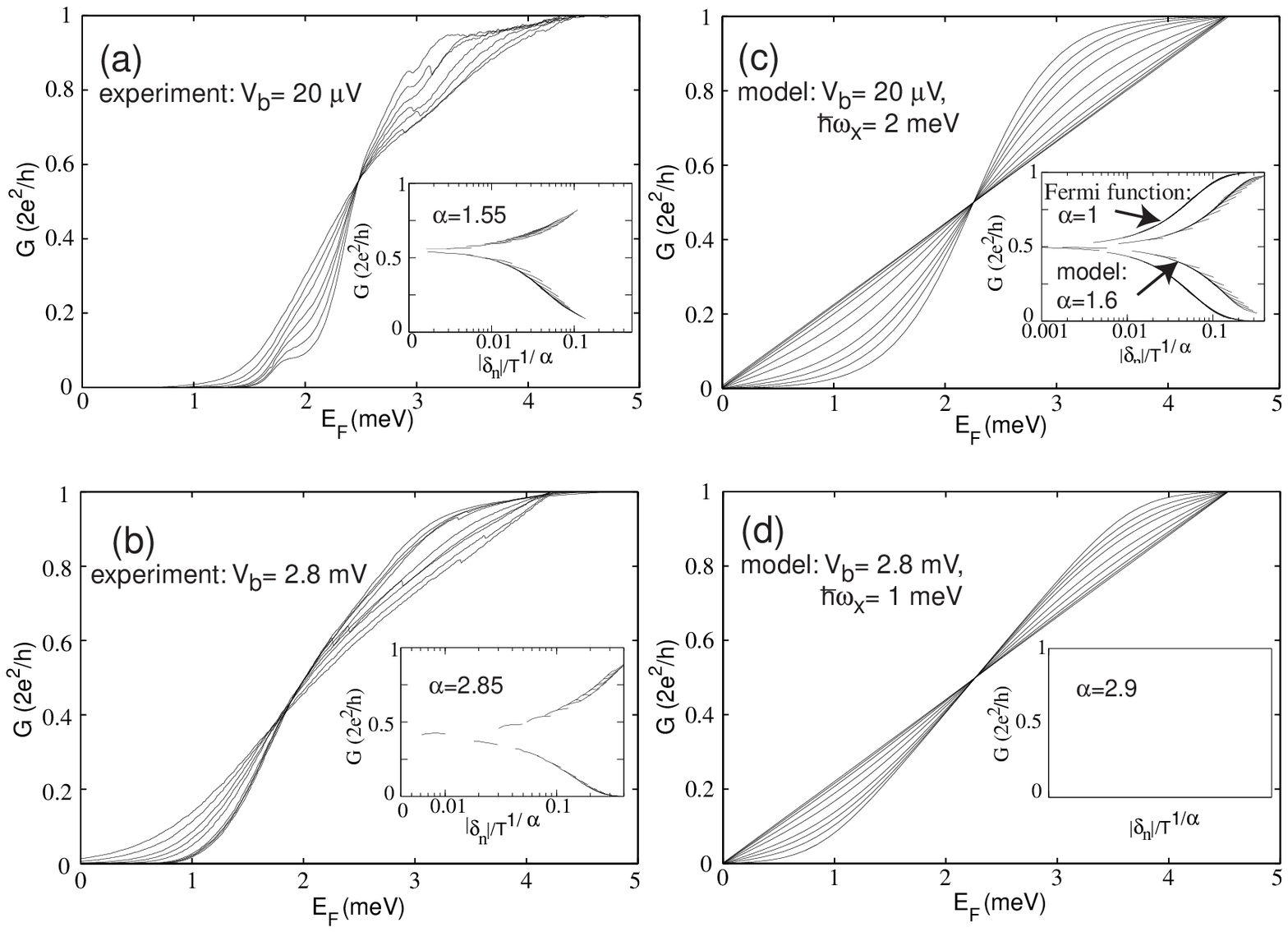}}
\caption{Conductance of the QPC as a function of the Fermi energy, 
as measured for $V_{sd}$ = 20 $\mu$V 
(a) and for $V_{sd}$ = 2.8 mV (b). The Fermi energy was calculated from $V_{tg}$
using the lever arm (see text). The insets show the corresponding 
scaling plots and the scaling parameters obtained. In (c) and (d), the calculated conductance within 
the model described in the text for the corresponding $V_{sd}$ are shown, 
together with the scaling functions of the calculated conductance 
traces. In addition, in the inset in (c), the scaled thermally smeared 
Fermi function is shown for comparison, which has $\alpha$ =1.   }
\label{Fig2}
\end{figure}

\begin{multicols}{2}
where $T(\epsilon)$ is given by eq. (2), and use $\omega_{x}$ as a parameter 
in order to obtain agreement with the 
experiment (Figs. 2c and 2d). With eq. (3), we can describe the 
experimental results, provided we allow $\omega_{x}$
to decrease as $V_{b}$ is increased. Such an effect is quite 
reasonable, since the potential drop in transport direction is 
superimposed to the saddle point potential, which may result in a reduction of the 
curvature $\omega_{x}$ at the saddle point. Furthermore, we have assumed in the 
model that $\omega_{y}$ is independent of $V_{b}$, as observed 
experimentally, i.e. the separation of the steps in $V_{tg}$ does not 
depend on $V_{b}$. Also, our model does not make any assumptions on how the voltage drops 
along the QPC.\cite{Ulreich1998}\\
The conductance traces obtained from eq. (3) no longer scale exactly onto a 
single curve (Figs. 2c, d). However, the scaling analysis not only allows 
us to express the behavior around the fixed point in terms of a single 
parameter, but also shows that the imperfect scaling of the 
experimental data can be expected from models differing from the 
oversimplified  one described by eq. (1).\\ 
The insets in Fig. 2 show the scaled data (scaling functions) of the 
main figures. As $V_{b}$ increases, $\omega_{x}$ 
has to be reduced in order to obtain agreement between the scaling 
parameters in the experiment and in the model. This behavior is more clearly shown in Fig. 3, where 
we plot the results of many simulations and compare them with the 
experimental data. It can be seen that for $eV_{b}\geq \hbar\omega_{x}$,
 the $\omega_{x}$ needed in the simulation in order to 
reproduce the experimental scaling parameter becomes smaller. The reason for this is possibly 
again the simplifications made in our model, or nonlinearities in the 
current-voltage characteristics, which are expected as soon as $e\cdot 
V_{b}$ gets of the order of the suband spacing $\hbar\omega_{y}$.
For small source-drain bias voltages, however, the experimentally obtained scaling 
parameter can be interpreted as a measure of $\omega_{x}$.\\
The  model developed by Meir in Ref.
\onlinecite{Meir99} describes the recently observed metal-insulator 
transition (MIT) in two dimensions in terms of a network of quantum point 
contacts with statistically distributed conductances. This network 
can be thought of being composed of two characteristic conductances, 
a metallic  one ($\sigma_{m}$), and an insulating one ($\sigma_{i}$), 
which determine the scaling behavior of the network around the 
percolation threshold. Thus, this model cannot be applied to the MIT at a \textit{single} 
QPC, as investigated in our experiment. Nevertheless, our results are related to Meir' s model, 
since they demonstrate that the shape of the saddle potential influences 
$\sigma_{m}$ and $\sigma_{i}$. 
Within a simple approximation, however, a connection between the MIT 
on a QPC and that one in a disordered 
two-dimensional carrier gas can be made. In the case of a disordered system 
consisting of conductive puddles, connected by \textit{identical} QPCs, 
scaling parameter obtained experimentally from scaling the MIT 
of the disordered system is directly related to $\omega_{x}$. 
It is the characterizing parameter for the connections between the 
puddles (note that within our simple model, the macroscopic voltage drop 
measured corresponds to the voltage drop across a single QPC, 
except for a geometry factor)
 \begin{figure}
\centerline{\epsfxsize=7.0cm \epsfbox{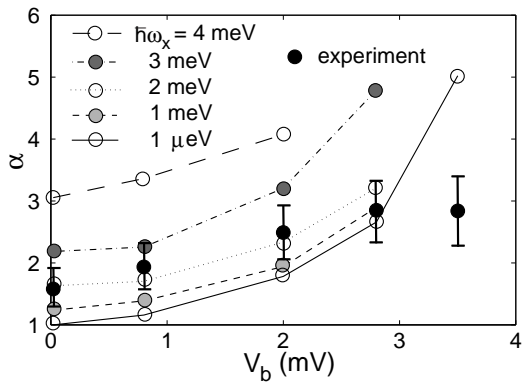}}
\caption{Comparison between the experimentally obtained values for 
$\alpha$ (full circles) and those obtained for various 
$\omega_{x}$ (see legend) within our model. The model 
predicts an increasing scaling parameter as $V_{sd}$ increases, which 
is also seen in the experiment. 
For V$_{b}\leq $ 2 mV, the experimental numbers for $\alpha$ agree 
well with the model, assuming $\hbar \omega_{x} \approx$ 2 meV. For 
V$_{b}> $ 2 mV, deviations occur (see text). }
\label{Fig3}
\end{figure}
The samples which perhaps come closest to such a scenario are those 
studied by Ribeiro et al. \cite{Ribeiro99} There, self-assembled 
quantum dots are embedded in the plane of the electron gas. The dots, 
which act as repulsive scatterers, have a diameter of only 30 nm,
a separation of $\approx$ 80 nm on average, and are very homogeneous 
in size. It is thus quite possible that between the self-assembled dots, QPCs are formed, 
in which only the lowest subband is occupied. Analyzing the MIT observed in these disordered 
samples gives $\alpha$ = 2.6 at small 
bias voltages, \cite{Ribeiro99} which, within our model, corresponds to a 
realistic value of $\hbar \omega_{x}\approx$ 3.5 meV.\\

In summary, we have observed a sequence of 
fixed points in a single quantum point contact. The scaling 
behavior of the temperature dependent conductance around the fixed 
point in the lowest subband has been studied, and significant deviations from simple Fermi function 
scaling have been found. Besides temperature, the scaling parameter
is also determined by the potential shape of the QPC in transport 
direction, which we have quantified by the curvature $\omega_{x}$, and by the source-drain voltage.
Our experiment and analysis suggests that in order to further check the 
validity  of Meir's model, an artificial, random network of quantum 
point contacts should be studied, which has become feasible by using 
lithography with an atomic force microscope.\\ 

Financial support by the Schweizerischer Nationalfonds is gratefully 
acknowledged.

\end{multicols}
\end{document}